\documentclass[
 reprint,
 nofootinbib,
superscriptaddress,
amsmath,
amssymb,
aps,
floatfix,
]{revtex4-2}

\usepackage{graphicx}

\usepackage{dcolumn}
\usepackage{bm}
\usepackage{color}
\usepackage[normalem]{ulem}

\usepackage[colorlinks,urlcolor=black,citecolor=black,linkcolor=black]{hyperref}

\begin{document}

\preprint{topo}


\title{Intrinsically chiral exciton polaritons in an atomically-thin semiconductor}


\author{M.~J.~Wurdack}
\email{mwurdack@stanford.edu}

\affiliation{Institute of Solid State Physics, Friedrich Schiller University Jena, 07743 Jena, Germany.}
\affiliation{Department of Quantum Science and Technology, Research School of Physics, The Australian National University, Canberra, ACT 2601, Australia.}
\affiliation{Institute of Applied Physics, Friedrich Schiller University Jena, 07745 Jena, Germany.}
\affiliation{Abbe Center of Photonics, Friedrich Schiller University Jena, 07745 Jena, Germany.}
\affiliation{Department of Chemical Engineering, Stanford University, Stanford, CA 94305, USA.}

\author{$^{,\kern 0.15em\dagger}$\ I.~Iorsh}
\thanks{These authors contributed equally.}
\affiliation{Department of Physics, Engineering Physics and Astronomy, Queen’s University, Kingston, Ontario, K7L 3N6, Canada.}

\author{S.~Vavreckova}

\affiliation{Institute of Applied Physics, Friedrich Schiller University Jena, 07745 Jena, Germany.}
\affiliation{Abbe Center of Photonics, Friedrich Schiller University Jena, 07745 Jena, Germany.}
\affiliation{Department of Quantum Science and Technology, Research School of Physics, The Australian National University, Canberra, ACT 2601, Australia}

\author{T.~Bucher}
\affiliation{Institute of Solid State Physics, Friedrich Schiller University Jena, 07743 Jena, Germany.}
\affiliation{Institute of Applied Physics, Friedrich Schiller University Jena, 07745 Jena, Germany.}
\affiliation{Abbe Center of Photonics, Friedrich Schiller University Jena, 07745 Jena, Germany.}

\author{M.~Król}
\affiliation{Department of Quantum Science and Technology, Research School of Physics, The Australian National University, Canberra, ACT 2601, Australia}

\author{Z.~Fedorova} 
\affiliation{Institute of Solid State Physics, Friedrich Schiller University Jena, 07743 Jena, Germany.}
\affiliation{Institute of Applied Physics, Friedrich Schiller University Jena, 07745 Jena, Germany.}
\affiliation{Abbe Center of Photonics, Friedrich Schiller University Jena, 07745 Jena, Germany.}

\author{E.~Estrecho}%
\affiliation{Department of Quantum Science and Technology, Research School of Physics, The Australian National University, Canberra, ACT 2601, Australia}

\author{S.~Klimmer}
\affiliation{Institute of Solid State Physics, Friedrich Schiller University Jena, 07743 Jena, Germany.}
\affiliation{Abbe Center of Photonics, Friedrich Schiller University Jena, 07743 Jena, Germany.}

\author{L. P. L. Mawlong}
\affiliation{Manufacturing, CSIRO, West Lindfield, Sydney, NSW, 2070 Australia.}

\author{H.~Deng}
\affiliation{Harbin Institute of Technology, Shenzhen 518055, China.}

\author{Q.~Song}
\affiliation{Harbin Institute of Technology, Shenzhen 518055, China.}

\author{T.~van der Laan}
\affiliation{Manufacturing, CSIRO, West Lindfield, Sydney, NSW, 2070 Australia.}

\author{G.~Soavi}
\affiliation{Institute of Solid State Physics, Friedrich Schiller University Jena, 07743 Jena, Germany.}
\affiliation{Abbe Center of Photonics, Friedrich Schiller University Jena, 07743 Jena, Germany.}

\author{T.~Pertsch}
\affiliation{Institute of Applied Physics, Friedrich Schiller University Jena, 07745 Jena, Germany.}
\affiliation{Abbe Center of Photonics, Friedrich Schiller University Jena, 07745 Jena, Germany.}

\author{F.~Eilenberger}
\affiliation{Institute of Applied Physics, Friedrich Schiller University Jena, 07745 Jena, Germany.}
\affiliation{Abbe Center of Photonics, Friedrich Schiller University Jena, 07745 Jena, Germany.}
\affiliation{Applied Optics and Precision Engineering IOF, Albert-Einstein-Str. 7, 07745 Jena, Germany}

\author{I.~Staude}
\affiliation{Institute of Solid State Physics, Friedrich Schiller University Jena, 07743 Jena, Germany.}
\affiliation{Institute of Applied Physics, Friedrich Schiller University Jena, 07745 Jena, Germany.}
\affiliation{Abbe Center of Photonics, Friedrich Schiller University Jena, 07745 Jena, Germany.}

\author{Y.~Kivshar}
\email{yuri.kivshar@anu.edu.au}
\affiliation{Nonlinear Physics Center, Research School of Physics, Australian National University, Canberra ACT 2601, Australia.}

\author{E.~A.~Ostrovskaya}
\email{elena.ostrovskaya@anu.edu.au}
\affiliation{Department of Quantum Science and Technology, Research School of Physics, The Australian National University, Canberra, ACT 2601, Australia}

\maketitle 

\textbf{Photonic bound states in the continuum (BICs) have emerged as a versatile tool for enhancing light-matter interactions by strongly confining light fields. Chiral BICs are photonic resonances with a high degree of circular polarisation, which hold great promise for spin-selective applications in quantum optics and nanophotonics. Here, we demonstrate a novel application of a chiral BIC for inducing strong coupling between the circularly polarised photons and spin-polarised (valley) excitons (bound electron-hole pairs) in atomically-thin transition metal dichalcogenide crystals (TMDCs). By placing monolayer WS$_2$ onto the BIC-hosting metasurface, we observe the formation of intrinsically chiral, valley-selective exciton polaritons, evidenced by circularly polarised photoluminescence (PL) at two distinct energy levels. The PL intensity and degree of circular polarisation of polaritons exceed those of uncoupled excitons in our structure by an order of magnitude. Our microscopic model shows that this enhancement is due to folding of the Brillouin zone creating a direct emission path for high-momenta polaritonic states far outside the light cone, thereby providing a shortcut to thermalisation (energy relaxation) and suppressing depolarisation. Moreover, while the polarisation of the upper polariton is determined by the valley excitons, the lower polariton behaves like an intrinsic chiral emitter with its polarisation fixed by the BIC. Therefore, the spin alignment of the upper and lower polaritons ($\uparrow\downarrow$ and $\uparrow \uparrow$) can be controlled by $\sigma^+$ and $\sigma^-$ polarised optical excitation, respectively. Our work introduces a new type of chiral light-matter quasi-particles in atomically-thin semiconductors and provides an insight into their energy relaxation dynamics.}

\begin{figure*} [htp!]
\includegraphics[width=\linewidth]{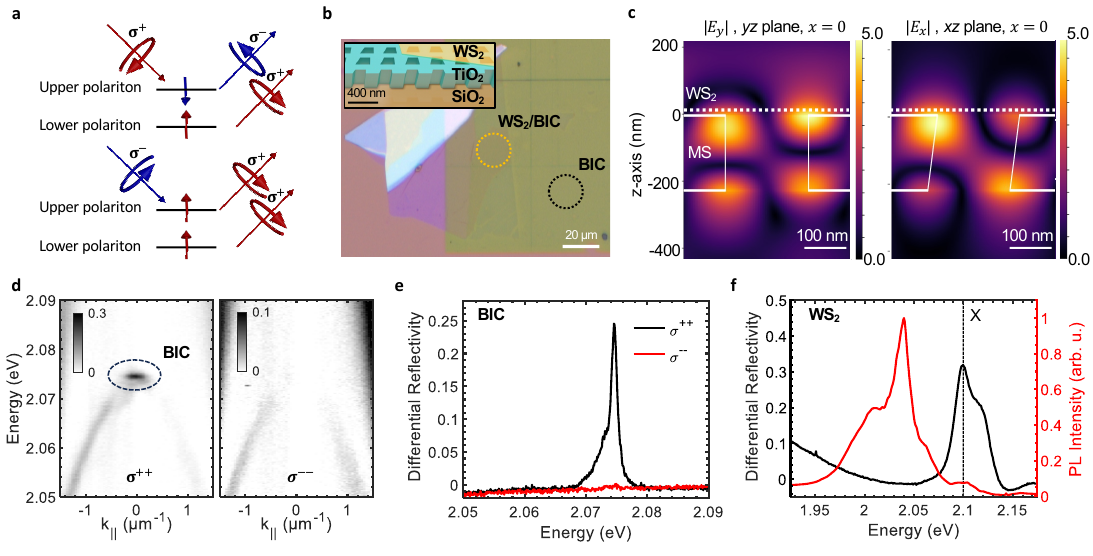}
\caption{ {\bf Optical properties of the WS$_2$/BIC heterostructure.}  {\bf a} Schematics of the spin configuration and emission properties of upper and  lower polaritons under (top) $\sigma^+$  and (bottom) $\sigma^-$ polarised optical excitation. {\bf b} Microscope image and (inset) schematic illustration of the heterostructure consisting of the metasurface governed by chiral BICs \cite{Chen2023} and the monolayer WS$_2$. The measured areas are marked with yellow and black circles. {\bf c} Electric field distribution of the metasurface at the BIC resonance $\lambda = 597~\mathrm{nm}$ in its linear polarisation bases (left) $\left|E_y\right|$ and (right) $\left|E_x\right|$ in the yz- and xz-plane, respectively.  {\bf d} Angle-resolved differential reflectivity spectra of the metasurface next to the position of the monolayer (black circle in {\bf b}) of (left) $\sigma^+$ and (right) $\sigma^-$ polarised white light, where the reflected light is filtered in its $\sigma^+$ and $\sigma^-$ polarisaton bases, respectively. The response of the chiral BIC is marked with a black dashed circle. {\bf e} Polarisation resolved reflectivity spectra of panel {\bf d} plotted along $k=0$. {\bf f} Reflectivity and photoluminescence spectrum at the position of the monolayer WS$_2$ (yellow circle in   {\bf b})  at $T\approx4~\mathrm{K}$, measured at large momenta ($k_{||}\approx 4.5~\mu\mathrm{m^{-1}}$, see Methods) to avoid the influence of the BIC.
}
\label{fig1}
\end{figure*}

Chiral photonic and electronic states are the building blocks for spin-based applications, which include quantum computing and communication, spintronics, and spin-selective sensing. Time-reversal symmetry breaking, parity-time-symmetry breaking or band-gap inversion can induce such chiral states \cite{Konig2007, Xiao2012, Mancini2015,Chen2023, Zhang2022, gao2018chiral, su2021direct}. Breaking the symmetry of a metasurface hosting photonic bound states in the continuum (BICs) \cite{Marinica2008,Koshelev2018,Kuhner2023} can generate chiral BICs that interact with either $\sigma^+$ or $\sigma^-$ polarised light, i.e., by separating their $\sigma^+$ and $\sigma^-$ polarisation bases in parameter space \cite{Zhang2022,Chen2023,Lv2024}. Similarly, in semiconducting atomically thin (monolayer) transition-metal dichalcogenide crystals (TMDCs), broken inversion symmetry combined with strong spin-orbit coupling leads to energetically degenerate band edges of the Brillouin zone with alternating spins, where spin-polarised bound electron-hole pairs (excitons) form in the K and K' valleys (see Fig. 1a) \cite{Xiao2012, Yu2015, Wang2018}. Thus, $\sigma^+$ and $\sigma^-$ polarised light can selectively probe K and K' valley excitons, respectively, which is facilitated by their high oscillator strength and strong photoluminescence (PL) \cite{Mak2010, Zeng2012, Chernikov2014, Wang2018, Gupta2023}. However, experimentally measured degrees of circular polarisation and valley coherence in TMDCs are often much lower than $1$ \cite{Zeng2012}, mostly due to inelastic scattering events, e.g., with phonons and disorder \cite{Gupta2023}, which strongly limits their potential applications. Scattering can be suppressed by forming exciton polaritons (polaritons herein) in the strong light-matter coupling regime \cite{Wurdack2021}, i.e., by placing a TMDC monolayer in a strongly confined light field of a microcavity or a photonic crystal \cite{Schneider2018}.   While polaritons in monolayer TMDCs are well studied \cite{Schneider2018,Luo2024}, these studies are limited to achiral photonic structures, where spin-selective behavior is mainly inherited by the valley-degenerate excitons \cite{Chen2017}.

Here, we demonstrate the formation of intrinsically chiral, valley-selective polaritons in a monolayer WS$_2$ placed on a metasurface governed  by chiral BICs \cite{Chen2023}. BICs were recently introduced as versatile photonic resonances for achieving polariton formation in a range of materials \cite{Ardizzone2022, Maggiolini2023, Wu2024}. The choice of material in this work, monolayer WS$_2$, is motivated by its spin-valley physics \cite{Plechinger2016} and the ability to reache strong light-matter coupling regime with its excitons in a wide temperature range, including room temperature \cite{Flatten2016, Zhang2018, Wurdack2021, Zhao2022, Maggiolini2023}. We perform experimental and theoretical reflectivity and PL studies showing that, in the strong light-matter coupling regime, diffraction overcomes thermalisation-induced losses and depolarisation, which leads to massively enhanced circularly polarised PL of both the upper and lower polariton branches. Remarkably, lower and upper polaritons can have parallel ($\uparrow\uparrow$) or anti-parallel ($\uparrow\downarrow$) spin-configurations, which can be switched with $\sigma^-$ or $\sigma^+$ polarised excitation, respectively, leading to co- or anti-polarised PL from the two energy levels (see Fig. \ref{fig1}a). This is because both photonic and excitonic constituents of the polaritons contribute to the spin-configurations, with the photonic (excitonic) contribution being stronger for the lower (upper) polariton. Thus, the lower polariton inherits most of its properties from the chiral BIC, and only interacts with and emits $\sigma^+$ polarised light \cite{Chen2023}. In contrast, the upper polariton shows rotating dipole behavior as inherited from the exciton, reversing the polarisation of the exciting light \cite{Bucher2024}. Additionally, the spin degree of freedom of valley excitons also manifests itself in the lower polariton, where the PL is the strongest when the exciton spin is aligned with that of the BIC. As such, combining the spin-selective light and matter states in a hybrid polaritonic state offers a novel pathway for the control and manipulation of chiral light emission.

\begin{figure*}[htp!]
\centering
\includegraphics[width=\textwidth]{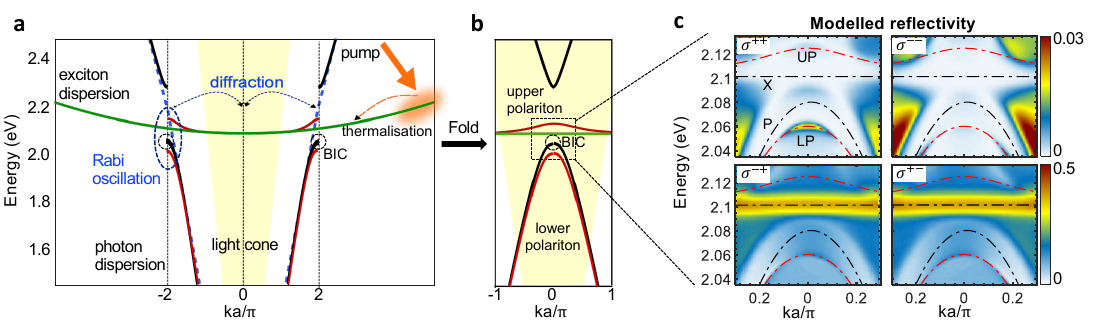}
\caption{{\bf Model of exciton-photon coupling.} {\bf a} Band structure of the (black) relevant photonic, (green) excitonic and (red) polaritonic modes of the sample in the strong exciton-photon coupling regime, and schematic illustration of relaxation and diffraction processes that lead to photoluminescence. {\bf b} Folded Brillouin zone of the photonic components, showing that both (green) excitons and (red) polaritons coexist within the light cone. The light cone of our experimental setup (see Methods) is shaded in yellow  and the location of the relevant chiral BIC ($\sigma^+$) on the photonic dispersion is marked with black dashed circles in panels a,b.  {\bf c} Modeled white light reflectivity in the far field, which is dominated by the polariton response (red dashed line) at the native polarisation of the chiral BIC ($\sigma^+$), and by the exciton response (X) when the polarisation of the reflected light is opposite to the polarisation of the incoming light (i.e., $\sigma^{+-/-+}$). The polariton (UP/LP) and uncoupled exciton/photon (X/P) dispersions are marked as red and black dashed lines, respectively.   }
\label{fig2}
\end{figure*}
\section*{Results}  

The structure investigated in this work consists of an exfoliated monolayer WS$_2$ placed on a slant-perturbed metasurface, i.e., a TiO$_2$ layer deposited on SiO$_2$ and patterned with a square array of trapezoid nanoholes (see Fig. \ref{fig1}b, {\color{black} SI} and Methods) \cite{Chen2023}. Good spatial overlap between the photonic near-field mode of the metasurface and the monolayer, confirmed by the calculation of the electric field distribution (see Fig. \ref{fig1}c and SI), enables the coupling between the photonic mode and the exciton in the monolayer. The sample is illuminated by either a polarised tungsten-halogen white light source or a continuous-wave (CW) laser tuned to the WS$_2$ bandgap and its reflectivity and PL properties are studied in the reflection geometry at cryogenic temperatures ($T\approx 4~\mathrm{K}$) (see Methods). {\color{black} Area-dependent reflectivity measurements show that, on the spatial scales relevant to our white light spot and monolayer size, the hole count and fabrication accuracy are sufficient to provide a high-Q resonance (see SI).
  }

The measured angle-resolved reflectivity spectrum next to the position of the monolayer (see Fig. \ref{fig1}b) shows a discrete mode, which only interacts with  $\sigma^+$ polarised white light at $k_{||} = 0$ and $E \approx 2.075~\mathrm{eV}$ (see Fig. \ref{fig1}d,e). In, particular, this mode disappears when the sample is illuminated by $\sigma^-$ polarised light.  This feature marks the optical response of the highly polarisation-selective chiral BIC with a Q-factor of $\sim1500$. Henceforth, $\sigma^{ir}$ convention indicates polarisations of the incoming ($i$) and reflected ($r$) light. The optical response of the monolayer was characterised via white light reflectivity and PL spectroscopy at large momenta ($k_{||}\approx 4.5~\mu\mathrm{m}^{-1}$) to avoid the influence of the BIC signal prevalent at $k_{||}=0$ (see Methods). The reflectivity and PL spectra in Fig. \ref{fig1}f show strong excitonic (X) absorption and weak PL emission at $E\approx 2.1~\mathrm{eV}$ \cite{Chernikov2014}, respectively. While at liquid Helium temperatures the PL of dark-type monolayer WS$_2$ mostly stems from charged and multi-body complexes, e.g., trions, biexcitons and charged biexcitons, which dominate the PL spectrum at lower energies between $\sim1.95~\mathrm{eV}$ and $2.08~\mathrm{eV}$ \cite{Mueller2018,Conway2022,Chen2018,Nagler2018,Singh2016,Plechinger2016,Ye2018,You2015}, the prominent exciton resonance at $\sim2.1~\mathrm{eV}$ in the reflectivity spectrum \cite{Chernikov2014}, which coincides with the weak exciton PL peak at the same energy, enables strong light-matter interactions and formation of exciton polaritons in this material \cite{Zhao2022}.

\begin{figure*}
\centering
\includegraphics[width=\textwidth]{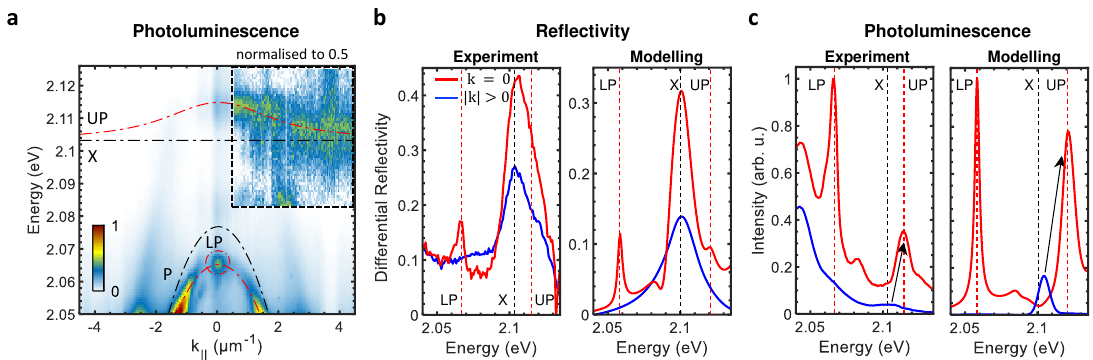}
\caption{{\bf Reflectivity and photoluminescence spectra.} {\bf a} Experimental angle-resolved photoluminescence spectrum, with  the  fitted (LP) lower and (UP) upper polariton branches (red dashed lines), and  the uncoupled (X) exciton and (P) photon dispersions (black dashed lines). The part of the upper polariton dispersion inheriting the chiral BIC properties is marked with a red dashed circle. The data in the inset are normalised to 0.5 for each $k_{||}$ value to visualise the peak positions of the low-intensity upper polariton emission.  {\bf b, c} Experimental and theoretical reflectivity and photoluminescence spectra at (red) $k_{||}=0$ nd (blue) $k_{||}\approx4.5~\mu\mathrm{m}^{-1}$, respectively. The arrows in panel (c) illustrate the order of magnitude enhancement of the PL intensities of the upper polariton between the lower energy state at $k_{||}\approx4.5~\mu\mathrm{m}^{-1}$ and the higher energy state at $k_{||}=0$ .}
\label{fig3}
\end{figure*}

To describe the salient features of the optical response in the strong light-matter coupling regime, we calculate the dispersion curves of the relevant photonic, excitonic and polaritonic modes (see Fig. \ref{fig2}a,b). The periodic structure of the metasurface creates a photonic band gap at $ka/\pi = \pm2$ (here $k_{||}\approx \pm18.2~\mu\mathrm{m^{-1}}$ with $a=340~\mathrm{nm}$, see Methods). The BIC is located at the energy maximum of the lower band, which is close to the exciton energy in monolayer WS$_2$. With sufficiently large energy-exchange interactions, i.e., Rabi oscillation frequencies, level repulsion occurs leading to formation of the upper and the lower polariton branches (see red lines in Fig. \ref{fig2}a). Folding of the Brillouin zone of the metasurface projects the dispersive branches from high-momentum states far outside the light cone of our experimental setup to low-momentum states, e.g., $ka/\pi =\pm2$ to $ka/\pi=0$, making the polariton branches visible within the light cone (see Fig. \ref{fig2}b). As such, both excitons (unfolded) and polariton branches (folded) coexist within the light cone. This is in contrast to polaritons in optical microcavities, where in the strong exciton-photon coupling regime the uncoupled excitons are only present far outside of the light cone \cite{Deng2010, Byrnes2014}.

When calculating the optical response (reflectivity) of the sample illuminated by circularly polarised white light (see Methods and Fig. \ref{fig2}c), we find that the lower polariton behaves similarly to the chiral BIC mode, with the reflectivity signal present at $\sigma^{++}$ and vanishing at $\sigma^{--}$ (cf. Figs. \ref{fig1}d and \ref{fig2}c, top panels). In contrast, the excitons, which are modeled as rotating dipoles, are responsible for the strong reflected signal with the polarisation opposite to that of the illumination $\sigma^{+-/-+}$, which is similar to light reflection at a metallic surface \cite{Scuri2018, Bucher2024}. This can be explained with our band structure (see Fig. \ref{fig2}a), which shows coexistence of excitons and polaritons within the light cone.

To verify our model experimentally, we performed angle-resolved white light reflectivity and PL spectroscopy at the position of the monolayer on the sample (see Fig. \ref{fig1}b). This was done by illuminating the sample with either a white light source or with a 561 nm (2.21 eV) cw-laser, which lies within the band gap of WS$_2$ ($E_g\approx2.4~\mathrm{eV}$) and quasi-resonantly excites excitons at large momenta. Figure \ref{fig3}a shows the angle-resolved PL intensity of the structure with the peak energies well reproduced by two coupled oscillator model describing the lower/upper polariton eigenvalues (LP/UP) emerging in the system of a photon (P) strongly coupled to an exciton (X): $E_{\rm LP/UP} = \frac{1}{2}[E_{\rm X} + E_{\rm P}\pm \sqrt{(2\hbar\Omega)^2 + \delta^2}]$, with the exciton-photon energy detuning $\delta=E_{\rm X} - E_{\rm P}=26.4~\mathrm{meV}$, and the Rabi splitting $2\hbar\Omega=42.0~\mathrm{meV}$. The excitonic ($|X|^2$) and photonic fractions ($|C|^2$) of polaritons can be described with the corresponding Hopfield coefficients: $|X_{\rm UP/LP}|^2=\frac{1}{2}\left[1\pm\delta/\sqrt{\delta^2+4\hbar^2\Omega^2}\right]$, $|C_{\rm UP/LP}|^2 = 1- |X_{\rm UP/LP}|^2$ \cite{Deng2010}, which in our system amount to $|X_{\rm UP/LP}|^2 \approx 0.76/0.24$ at $k_{||}=0$. The localised emission peak at $k_{||}=0$ and $E \approx 2.066~\mathrm{eV}$, reminiscent of the chiral BIC response at $E \approx 2.077~\mathrm{eV}$(see Fig. \ref{fig1}d), originates from the more photonic LP. The emission above 2.1 eV corresponds to the more excitonic UP. The UP disperison near $k_{||}=0$, when folded as shown in Fig. \ref{fig2}a,b,  is strongly modified by level repulsion in proximity to the BIC state causing anti-crossing with the exciton level (X). This confirms the strong coupling regime between the WS$_2$ excitons and the chiral photonic BIC. 

The PL signal, which is dominated by polaritonic states, is in stark contrast to the reflectivity spectrum (see Fig. \ref{fig3}b), which, as predicted by our model (see Fig. \ref{fig2}c), is strongly dominated by the exciton mode. Both high- and low-momenta spectra are well reproduced by our calculations (see polarisation-resolved spectra in the SI), with the sharp LP peak only present at low momenta (around $k=0$). This agreement confirms the accuracy of our model, and that both excitons and polaritons coexist within the light cone.

\begin{figure*}[htp]
\centering
\includegraphics[width=\textwidth]{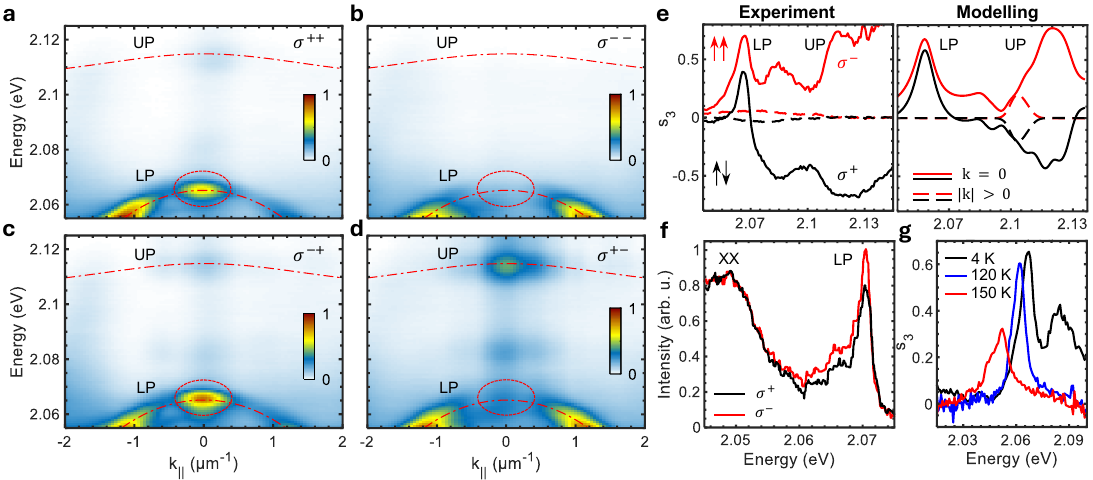}
\caption{{\bf Angle- and polarisation-resolved PL spectra.} {\bf a-d} Circularly polarised components of the angle-resolved PL spectra of the sample excited with a circularly polarised laser pump. The $\sigma^{pe}$ convention reflects the sign of the circular polarisation of the pump ($p$) and the PL emission ($e$).  {\bf e} Measured and modeled s$_3$ spectra with (black) $\sigma^-$ and (red) $\sigma^+$ polarised pump. The arrows ($\uparrow$,$\downarrow$) mark the spin configurations of the lower and upper polaritons, respectively, under (red) $\sigma^-$ and (black) $\sigma^+$ polarised excitation. {\bf f} PL spectrum of the LP with (black) $\sigma^+$ and (red) $\sigma^-$ polarised pump, with the PL spectra of the biexciton (XX) plotted as reference. {\bf g}  s$_3$ spectra of the LP at (black) $T=4~\mathrm{K}$ and (red) $T=150~\mathrm{K}$, respectively.  }
\label{fig4}
\end{figure*}

Remarkably, the exciton peak in the PL is strongly suppressed (see Fig. \ref{fig3}a,c). In particular, the PL intensity of polaritons at $k_{||}=0$ is an order of magnitude larger than that of the energetically lower more excitonic emission at $k_{||}\approx4.5 ~\mu\mathrm{m}^{-1}$ (see Fig. \ref{fig3}c). As thermalisation favors low energy states, the massively enhanced high energy emission contradicts thermalisation as main source of emission at $k_{||}=0$. To model this behavior, we further leveraged the Green's function approach (see Methods) for calculating the PL emission spectra using the random source approximation ~\cite{deych2007exciton} (see SI). In the calculation, we used the momentum distribution of the excitons PL as a fitting parameter to fit the experimentally measured PL spectra. Our model, which shows good agreement with our experimentally obtained data (see Fig. \ref{fig3}b,c) strongly suggests that diffraction is the main source of the polariton PL at $k=0$, as schematically shown in Fig. \ref{fig2}a. In particular, polaritons contribute to PL emission within the light cone due to the folding of the Brillouine zone of the metasurface, instead of thermalisation, leading to the observed PL enhancement. The excitons, however, can only relax into the light cone via thermalisation, i.e., inelastic scattering with phonons, which is less efficient. Our findings indicate that this PL mechanism is applicable to polaritons in periodic photonic structures, in general, where diffraction into the light cone can outpace energy relaxation to low momenta states, boosting their quantum yield. {\color{black} The additional low-energy peaks in the experimentally measured spectrum (Fig. \ref{fig3}c) arise from the charged and multi-body complexes, as described above. As these excitonic species do not possess sufficient oscillator strength to coherently couple to the resonant photonic mode in the metasurface, they are not accounted for in the theory.}

Finally, we investigate the polarisation properties of the polaritons by performing polarisation-resolved PL measurements.  Fig. \ref{fig4}a-d shows the circularly polarised components of the momentum-resolved PL spectra ($\sigma^{i+}$ in Fig. \ref{fig4}a,c and $\sigma^{i-}$ in Fig. \ref{fig4}b,d) upon excitation with $\sigma^+$ (Fig. \ref{fig4}a,d) or $\sigma^-$ (Fig. \ref{fig4}b,c) polarised light. Remarkably, at $k=0$, the LP emission is $\sigma^+$ polarised (Fig. \ref{fig4}a,c), following the behavior of the chiral BIC as seen in Fig. \ref{fig1}d. The polarisation of the UP, however, is opposite to that of the pump, both in $\sigma^{+-}$ and $\sigma^{-+}$ configurations, which is reminiscent of the rotating dipole behavior of the exciton. This is because the LP ($|C|^2 = 0.76$) mostly inherits the properties of the chiral photonic BIC state, while the UP ($|X|^2 = 0.76$) behaves more like a valley exciton. Consequently, the emitted light of the UP and LP have opposite(same) signs of circular polarisations at $k=0$ under $\sigma^+$($\sigma^-$) polarised excitation.
 
We further analyse the antisymmetric polarisation behavior by calculating the $s_3 = (I^+ - I^-)/(I^+ + I^-)$ spectra for both circular polarisations ($\sigma^+$ and $\sigma^-$) of the pump (see Fig. \ref{fig4}e). Here, UP and LP reach values of $s_3> +0.7$ with $\sigma^-$ polarised excitation, and $s_3\approx -0.7$ for the UP and $s_3\approx +0.4$ for the LP  with $\sigma^+$ polarised excitation. Interestingly, $s_3$ completely vanishes at large momenta, where the emitting states are mostly excitonic. This strong suppression of $s_3$ is  likely a result of inelastic scattering processes required for relaxation/thermalisation of excitons into the light cone. {\color{black} Additional inhomogeneous broadening due to the holes of the metasurface leads to a much larger experimental linewidth compared to our calculations, where only homogeneous linewidth broadening is considered.} 

At $k=0$, the strong enhancement of circular polarisation is well reproduced by our model, indicating that, in addition to boosting PL intensity, diffraction also strongly enhances polarisation of the polaritons by bypassing the inelastic scattering processes. In combination with the contribution from the intrinsically chiral BIC, it leads to the strong antisymmetric polarisation of the UP and LP, where changing the polarisation of the pump laser triggers switching between parallel and anti-parallel spin configurations of the two polariton states (see Fig. \ref{fig1}a). 

The role of the matter component (the valley exciton) in the chiral BIC-like lower polariton becomes evident when measuring the absolute intensity of its PL under circularly polarised excitation. We find that the quantum yield is the strongest when the spin of the valley exciton is aligned with the polarisation of the chiral BIC (see Fig. \ref{fig4}f). The reference spectrum produced by the uncoupled biexciton (XX) \cite{Nagler2018}, on the other hand, shows no enhancement. Further, the matter-component of the LP enables effective interaction with phonons, as shown in the temperature-dependent measurements (see Fig. \ref{fig4}g {\color{black} and SI}), where both the energy and degree of circular polarisation drop with increasing phonon energies. These results  demonstrate that by strongly coupling chiral BIC photons with valley excitons, we successfully created an intrinsically chiral hybrid state that possesses valley physics and effectively interacts with the surrounding medium, further distinguishing our findings from previous investigations of weakly coupled excitons \cite{Zhang2022}. {\color{black} Finally, we estimate a lifetime of 500 fs for the intrinsically chiral lower polaritons based on the lifetime and power-dependent measurements presented in the SI.}

In summary, by strongly coupling valley excitons with chiral photonic BICs, we have created a new type of chiral hybrid quasi-particles, whose two energy eigenstates can possess distinct spin configurations, either parallel ($\uparrow\uparrow$) or anti-parallel ($\uparrow \downarrow$), depending on the circular polarisation of the optical excitation source (see Fig. \ref{fig1}a). By combining the unique physics of a chiral photonic and a spin-valley electronic system, we therefore complemented the portfolio of spin-selective chiral states with a light-matter hybrid particle with optically controllable spin properties. The potential for ultrafast optical switching between $\uparrow\uparrow$ and $\uparrow\downarrow$ polarised emission in response to polarised light can be useful for spintronics, spin-based transistors, and chiral sensors.

\vspace{0.5cm}
\noindent\textbf{Methods:}
To make the metasurface, we deposited a 220 nm TiO$_2$ layer on a glass (SiO$_2$) substrate by electron beam evaporation (0.065 nm/s). The TiO$_2$ layer was then patterned using electron-beam lithography. This was done by depositing 20 nm of Cr and spin-coating a 80 nm thick film of PMMA. The PMMA was patterned using electron beam lithography, and the pattern then etched into the Cr layer with inductively coupled Cl$_2$ and O$_2$ plasma. Further, we etched the TiO$_2$ layer in a slant-etching system (as developed in \cite{Chen2023}) with reactive O$_2$, SF$_6$, Ar and CHF$_3$ ions. With this process, we created the slant-perturbed TiO$_2$ metasurface consisting of a square array of trapezoid nanoholes, with the parameters:  unit cell size $a=340~\mathrm{nm}$, width of hole $w=210~\mathrm{nm}$, height of hole $h=220~\mathrm{nm}$, slant angle $\phi = 0.1$ and in-plane deformation angle $\alpha = 0.12$, as presented in \cite{Chen2023} and schematically shown in Fig. \ref{fig1}b. The monolayer WS$_2$ was mechanically exfoliated from a bulk WS$_2$ crystal (sourced from HQ graphene) onto a gel-film (Gel Pak), and transferred onto the metasurface using the dry transfer technique.

The reflectivity of the sample was calculated using the rigorous coupled wave analysis (RCWA)~\cite{schlipf2021rigorous}. The RCWA method allows us to calculate both the angular and frequency spectra of the polarisation-resolved reflection coefficient and the total scattering matrix $S^{\mathbf{QQ'}}$ connecting the multiple diffraction orders of the incoming and outgoing waves $\mathbf{E}^{\mathbf{Q}}=S^{\mathbf{QQ'}}\mathbf{E}^{\mathbf{Q'}}$, where $\mathbf{Q}$ index labels different diffraction orders. The scattering matrix is also used to calculate the dyadic Green's  function of the metasurface, which we employ for calculating the PL emission spectra of our structure (see SI).

The PL and reflectivity measurements were conducted in a closed-cycle helium cryostat (Montana Instruments s50). For the PL measurements, the sample was excited with a 561 nm wavelength continuous-wave diode-pumped solid-state laser at a power of $30\mu W$. The source for the reflectivity measurements was a stabilised tungsten-halogen white light source. The laser or white light was focused with a high-NA objective (100x/0.9 NA) onto the sample surface, creating a light spot size of 0.45 $\mu$m in the focus of the objective. The reflected or emitted light from our structure was separated from the incoming light by a  non-polarising 30:70 beam splitter cube, and collected with an array of lenses in a 4-f arrangement, where the back focal plane was focused onto the spectrometer slit of an Andor Shamrock i750 spectrograph, equipped with an Andor iXon 897 Ultra EMCCD camera, for imaging the presented momentum-resolved spectra. By employing a spatial filter in the real space image of our 4f-setup, we ensured that only the light stemming from the marked areas in Fig. \ref{fig1}b is collected and analysed.  The presented reflectivity data were obtained by differential reflectance spectroscopy $\Delta R/R_{Ref} = (R - R_{Ref})/R_{Ref}$, using the reflectivity spectrum of the TiO$_2$ coated glass surface next to the metasurface as a reference (Ref). For the polarisation-resolved measurements, we prepared the polarisation of the excitation beam with a linear polariser and quarter-wave phase plate. The polarisation of the detected light was analysed by a combination of super-achromatic quarter-wave phase plate and a linear polariser. To increase the signal-to-noise ratio in the presented polarisation-resolved images (\ref{fig4}a-d), we performed a moving average calculation in momentum space, effectively applying the average of 9 neighboring pixels along $k_{||}$ to each pixel.

\vspace{0.5cm}

\noindent\textbf{Ethics and inclusion:} This study was conducted with a strong commitment to ethical and inclusive research practices.

\vspace{0.5cm}
\noindent\textbf{Data availability:}
The data that support the findings of this study are available from the corresponding authors upon reasonable request.

\vspace{0.5cm}

\noindent\textbf{Code availability:} The custom code used in this study is available from the corresponding author upon reasonable request.

\vspace{0.5cm}

\noindent\textbf{Acknowledgements:}
This work was funded by the Australian Research Council (grants CE170100039, DP210101292 and DE220100712), the Deutsche Forschungsgemeinschaft (DFG, German Research Foundation) - CRC/SFB 1375 NOA ”Nonlinear Optics down to Atomic scales” (Project number 398816777), the International Research Training Group 2675 “META-Active (project number 437527638) and the International Technology Center Indo-Pacific (ITC IPAC) via Army Research Office (contract FA520923C0023). M.W. acknowledges support by Schmidt Science Fellows, in partnership with Schmidt Sciences and Rhodes Trust. The authors thank Mikhail Glazov for fruitful discussions.


\bibliography{refs.bib}

\end{document}


\preprint{topo}


\title{Supplementary Information \\ Intrinsically chiral exciton polaritons in an atomically-thin semiconductor}


\author{M.~J.~Wurdack}
\email{mwurdack@stanford.edu}

\affiliation{Institute of Solid State Physics, Friedrich Schiller University Jena, 07743 Jena, Germany.}
\affiliation{Department of Quantum Science and Technology, Research School of Physics, The Australian National University, Canberra, ACT 2601, Australia.}
\affiliation{Institute of Applied Physics, Friedrich Schiller University Jena, 07745 Jena, Germany.}
\affiliation{Abbe Center of Photonics, Friedrich Schiller University Jena, 07745 Jena, Germany.}
\affiliation{Department of Chemical Engineering, Stanford University, Stanford, CA 94305, USA.}

\author{$^{,\kern 0.15em\dagger}$\ I.~Iorsh}
\thanks{These authors contributed equally.}
\affiliation{Department of Physics, Engineering Physics and Astronomy, Queen’s University, Kingston, Ontario, K7L 3N6, Canada.}

\author{S.~Vavreckova}

\affiliation{Institute of Applied Physics, Friedrich Schiller University Jena, 07745 Jena, Germany.}
\affiliation{Abbe Center of Photonics, Friedrich Schiller University Jena, 07745 Jena, Germany.}
\affiliation{Department of Quantum Science and Technology, Research School of Physics, The Australian National University, Canberra, ACT 2601, Australia}

\author{T.~Bucher}
\affiliation{Institute of Solid State Physics, Friedrich Schiller University Jena, 07743 Jena, Germany.}
\affiliation{Institute of Applied Physics, Friedrich Schiller University Jena, 07745 Jena, Germany.}
\affiliation{Abbe Center of Photonics, Friedrich Schiller University Jena, 07745 Jena, Germany.}

\author{M.~Król}
\affiliation{Department of Quantum Science and Technology, Research School of Physics, The Australian National University, Canberra, ACT 2601, Australia}

\author{Z.~Fedorova} 
\affiliation{Institute of Solid State Physics, Friedrich Schiller University Jena, 07743 Jena, Germany.}
\affiliation{Institute of Applied Physics, Friedrich Schiller University Jena, 07745 Jena, Germany.}
\affiliation{Abbe Center of Photonics, Friedrich Schiller University Jena, 07745 Jena, Germany.}

\author{E.~Estrecho}%
\affiliation{Department of Quantum Science and Technology, Research School of Physics, The Australian National University, Canberra, ACT 2601, Australia}

\author{S.~Klimmer}
\affiliation{Institute of Solid State Physics, Friedrich Schiller University Jena, 07743 Jena, Germany.}
\affiliation{Abbe Center of Photonics, Friedrich Schiller University Jena, 07743 Jena, Germany.}

\author{L. P. L. Mawlong}
\affiliation{Manufacturing, CSIRO, West Lindfield, Sydney, NSW, 2070 Australia.}

\author{H.~Deng}
\affiliation{Harbin Institute of Technology, Shenzhen 518055, China.}

\author{Q.~Song}
\affiliation{Harbin Institute of Technology, Shenzhen 518055, China.}

\author{T.~van der Laan}
\affiliation{Manufacturing, CSIRO, West Lindfield, Sydney, NSW, 2070 Australia.}

\author{G.~Soavi}
\affiliation{Institute of Solid State Physics, Friedrich Schiller University Jena, 07743 Jena, Germany.}
\affiliation{Abbe Center of Photonics, Friedrich Schiller University Jena, 07743 Jena, Germany.}

\author{T.~Pertsch}
\affiliation{Institute of Applied Physics, Friedrich Schiller University Jena, 07745 Jena, Germany.}
\affiliation{Abbe Center of Photonics, Friedrich Schiller University Jena, 07745 Jena, Germany.}

\author{F.~Eilenberger}
\affiliation{Institute of Applied Physics, Friedrich Schiller University Jena, 07745 Jena, Germany.}
\affiliation{Abbe Center of Photonics, Friedrich Schiller University Jena, 07745 Jena, Germany.}
\affiliation{Applied Optics and Precision Engineering IOF, Albert-Einstein-Str. 7, 07745 Jena, Germany}

\author{I.~Staude}
\affiliation{Institute of Solid State Physics, Friedrich Schiller University Jena, 07743 Jena, Germany.}
\affiliation{Institute of Applied Physics, Friedrich Schiller University Jena, 07745 Jena, Germany.}
\affiliation{Abbe Center of Photonics, Friedrich Schiller University Jena, 07745 Jena, Germany.}

\author{Y.~Kivshar}
\email{yuri.kivshar@anu.edu.au}
\affiliation{Nonlinear Physics Center, Research School of Physics, Australian National University, Canberra ACT 2601, Australia.}

\author{E.~A.~Ostrovskaya}
\email{elena.ostrovskaya@anu.edu.au}
\affiliation{Department of Quantum Science and Technology, Research School of Physics, The Australian National University, Canberra, ACT 2601, Australia}

\maketitle 
\begin{figure}[htp]
\centering
\includegraphics[width=8.5 cm]{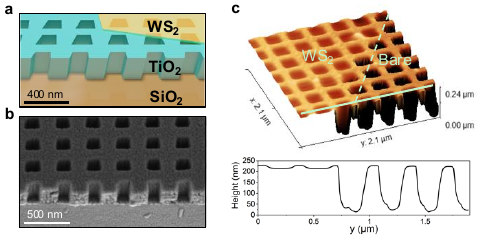}
\caption{{Design of the hybrid metasurface/2D semiconductor heterostructure.} {\bf a} Schematic illustration of the heterostructure. {\color{black} {\bf b} Scanning electron microscope (SEM) image of the metasurace. {\bf c} Atomic force microsce (AFM) image of the heterostructure. The interface between metasurface/WS$_2$ and bare metasurface is marked with a dashed white line, and the position of the (bottom) line profile with a solid white line.}}
\label{SI_fig1new}
\end{figure}
{\bf Sample Design {\color{black} and Characterisation}:} A schematic illustration, {\color{black} SEM image, and AFM image of the  bare slant-perturbed TiO$_2$ metasurface and metasurface/WS2$_2$ heterostructure are shown in Fig. \ref{SI_fig1new}a-c, respectively. As shown in the images, the} metasurface consists of a square array of trapezoid nanoholes, with designed unit cell size $a=340~\mathrm{nm}$, hole width $w=210~\mathrm{nm}$, hole height $h=220~\mathrm{nm}$, slant angle $\phi = 0.1$ and in-plane deformation angle $\alpha = 0.12$, as discussed more in detail in \cite{Chen2023}. {\color{black} The monolayer was mechanically exfoliated and transferred onto the metasurface using the dry transfer method. The AFM and SEM images show that the fabricated structure is consistent with the design, and that the arrangement and geometries of the holes are regular.}

\begin{figure}[htp]
\centering
\includegraphics[width=8.5 cm]{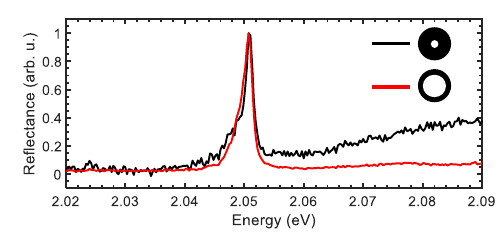}
\caption{{\color{black} High resolution spectra of the chiral photonic resonance averaged over an area of (black) $\sim400~\mu\mathrm{m}$ and (red) $\sim2000~ \mu\mathrm{m}^2$.}}
\label{SI_fig2_new}
\end{figure}

{\color{black}
To evaluate local fluctuations of the photonic resonance on the spatial scales relevant to our monolayer size, we compared the reflectivity spectrum of the chiral mode averaged over an area smaller than the monolayer ($\sim400~\mu\mathrm{m}^2$) to that of a larger area ($\sim2,000 \mu\mathrm{m}^2$), consisting of $\sim4,000$ and $\sim 20,000$ holes, respectively. This experiment was done by employing a spatial filter in the real space plane of our experimental setup and extracting  the momentum-space spectrum of the $\sigma^+$ component of the reflected white light at $k=0$. Figure \ref{SI_fig2_new} shows the high-resolution spectra of the chiral resonance for both configurations. When comparing the linewidths, we can observe that local fluctuations of the hole geometries across the metasurface lead to the inhomogeneous broadening of the resonance. However, since the amount of broadening is only in the order of 1 meV, which is an order of magnitude smaller than the inhomogeneous broadening of the excitons (see main text), we conclude that local fluctuations are relatively small and that the spatial geometries of the holes are sufficiently consistent for producing a high-Q resonance across the whole monolayer area.
}

{\bf Theory:} The metasurface is a structure with discrete translational invariance. The in-plane component of the wavevector is conserved modulo reciprocal lattice vectors $\mathbf{Q}$. In the systems with continuous translational invariance we can express the electric field vector $\mathbf{E}_{\mathbf{k}}$ as
\begin{align}
\mathbf{E}_{\mathbf{k}}(z) = \int dz' G_{\mathbf{k}}(z,z') \mathbf{P}_{\mathbf{k}}(z'),
\end{align}
where $\mathbf{k}$ is the in-plane wavevector, $\mathbf{P}$ is the polarisation at a specific wavevector and specific position along the $z$ axis. In the system with discrete translational invariance we instead should write
\begin{align}
\mathbf{E}_{\mathbf{k+Q}}(z) = \int dz' G_{\mathbf{k}}^{\mathbf{Q,Q'}}(z,z') \mathbf{P}_{\mathbf{k+Q'}}(z'), \label{E_init}
\end{align}
where $\mathbf{Q,Q'}$ span over all the reciprocal lattice vectors and the summation is taken over the repeating indices $\mathbf{Q'}$. For each specific values of $\mathbf{k},z,z'$, $G$ is the matrix consisting of $3\times 3$ blocks labeled by $\mathbf{Q}$ and $\mathbf{Q'}$ which  connect the 3 electric field components with momenta $\mathbf{k+Q}$ and at position $z$ to the polarisation with momenta $\mathbf{k+Q'}$ at position $z'$. In our analysis we will only consider in-plane polarisation induced by excitons and model in-plane components of the emitted electric field, therefore we will only consider $2\times 2$ blocks. 

We assume that the monolayer is placed at the $z=0$ plane and that both incident and emitted radiation is in the $z>0$ half space.  For further analysis we will need the Green's functions corresponding to $z=z'=0$ and to $z\rightarrow \infty , z'=0$. While the former connects the polarisation at the plane of the monolayer to the electric field at the same plane, the latter connects the monolayer polarisation to the far field which is ultimately detected in the experiment. 

\begin{figure}[htp]
\centering
\includegraphics[width=8.5 cm]{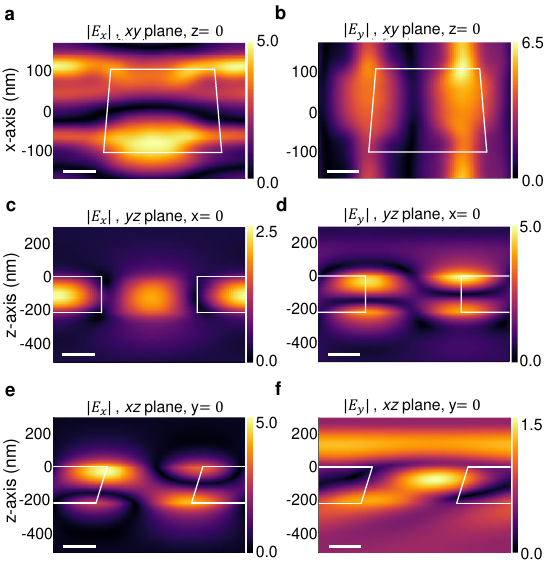}
\caption{{\bf Electric field distribution of the BIC mode at $\lambda= 597~\mathrm{nm}$.} {\bf a,b} $E_x$ and $E_y$ in the xy-plane at the top surface, where we place our monolyer (z=0), respectively.  {\bf c-d} $E_x$ and $E_y$ components of the cross-sections in the  (c-d) yz- and (e-f) xz-planes, at x=0 or y=0, respectively. The contour of the metasurface is marked as white solid lines. The scale bar size is 50 nm.}
\label{SI_fig1}
\end{figure}

The Green's function can be expressed via the reflection matrix $\hat{r}$:
\begin{align}
G^{\mathbf{QQ'}}_{\mathbf{k}}(0,0)= [\hat{I}+\hat{r}]\cdot[\delta_{\mathbf{QQ'}}\otimes G_{0,\mathbf{k+Q}}],
\end{align}
where $\hat{I}$ is the identity matrix, $G_0$ is the Green's function for the free space:
\begin{align}
G_{0,\mathbf{k+Q}}=\frac{i}{2k_0^2 k_z} \begin{pmatrix} k_0^2-k_x^2 & -k_x k_y \\ -k_x k_y & k_0^2-k_y^2 \end{pmatrix},
\end{align}
and $\hat{r}$ is the reflection consisting of $2\times 2$ blocks 
\begin{align}
\hat{r}^{\mathbf{QQ'}}=\begin{pmatrix}  r_{xx}^{\mathbf{QQ'}} & r_{xy}^{\mathbf{QQ'}} \\ r_{yx}^{\mathbf{QQ'}} & r_{yy}^{\mathbf{QQ'}} \end{pmatrix}
\end{align}
The reflection matrix is computed numerically using the rigorous coupled wave analysis (RCWA)~\cite{schlipf2021rigorous}. We note that RCWA allows us to calculate the generalised transfer matrices in the whole structure, which in turn can be used to restore the local field distributions. The electric field distribution at the wavelength of the chiral BIC $\lambda\approx=597~\mathrm{nm}$ maps in three orthogonal planes are shown in Fig.~\ref{SI_fig1}.

The polarisation of excitons $\mathbf{P}_{\mathbf{q}}(z,t)=\mathbf{P}_{\mathbf{q}}(t)\delta(z)$ satisfies the equation:
\begin{align}
\dot{\mathbf{P}}_{\mathbf{q}}(t)=&(-i\omega_X(\mathbf{q})-\Gamma-i\frac{1}{2}\mathbf{\Omega}_{\mathbf{q}}\mathbb{\sigma})\mathbf{P}_{\mathbf{q}}-\\ & \gamma \mathbf{E}_{\mathbf{q}}(z=0,t)+\mathbb{\chi}_{\mathbf{q}}(t), \label{Pol_eq}
\end{align}
where $\omega_X, \Gamma$ are the frequency and non-radiative decay rates of excitons, $\Omega_{\mathbf{q}}$ is the effective exciton magnetic field due to the TE-TM exciton splitting, $\gamma$ is the excitonic radiative decay rate, and $\chi$ is the pumping rate, i.e. the rate of exciton injection from the reservoir. The pumping rate $\chi$ is modeled within the random source model~\cite{deych2007exciton} as a random process satisfying
\begin{align}
\langle \chi_{\alpha}(\mathbf{q},t) \rangle =&0,\\ \quad \quad  \langle \chi_{\alpha}^{*} (\mathbf{q},t) \chi_{\beta} (\mathbf{q'}, t')\rangle = &\rho_{\alpha\beta}(\mathbf{q},t) \delta_{\mathbf{q,q'}} \delta(t-t') 
\end{align}
where $\rho_{ij}$ is defined both by the polarisation of the pump and by the relaxation kinetics of excitons due to exciton-phonon and exciton-exciton interactions.

\begin{figure}[htp]
\centering
\includegraphics[width=8.5 cm]{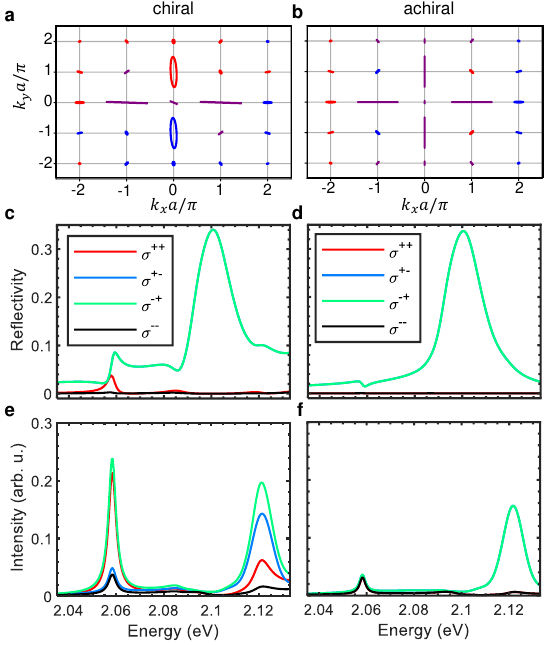}
\caption{{\bf Modeled polarisation properties, and reflectivity and PL spectra of exciton-polaritons in (with slant) chiral and (without slant) achiral geometries.} {\bf a,b} Polarisation of the lower polariton in the diffractions orders, with (blue) $\sigma^-$, (red) $\sigma^+$ and (purple) linear polarisation,  and the ellipsity corresponding to the degree of circular polarisation, for chiral and achiral structures, respectively. {\bf c,d} Polarisation resolved reflectivity spectra of the modeled chiral and achiral structures, respectively, where $\sigma^{ir}$ convention indicates polarisations of the incoming ($i$) and reflected ($r$) light. {\bf e,f} Polarisation resolved PL spectra of the modeled chiral and achiral structures, respectively, where $\sigma^{pe}$ convention indicates polarisations of the pump ($p$) and emitted light ($e$).
}
\label{SI_fig2}
\end{figure}

The equation~\eqref{Pol_eq} for $\mathbf{P}$ can be solved exactly, and then $\mathbf{P}$ can be substituted into Eq.~\eqref{E_init},  which results in the equation for the total field at $z=0$ which again can be solved exactly. Finally, we obtain the far-field expression for the electric field:   

\begin{align}
E_{\alpha}(\mathbf{k},z\rightarrow\infty, \omega) = \\G_{\mathbf{k},\alpha\beta}^{0\mathbf{Q}}(-\infty,0,\omega) [\hat{\alpha}_0(\mathbf{k+Q,\omega})^{-1}\\\delta_{\mathbf{QQ'}}-\hat{G}_{\mathbf{k}}(0,0,\omega)]^{-1,\mathbf{QQ'}}_{\beta\xi}\chi_{\xi}^{\mathbf{Q'}}/\gamma \label{El_full},
\end{align}
where $\alpha_0$ is the bare exciton polarisability, $\alpha_0^{-1}(\mathbf{q},\omega)=(\omega-\omega_X(\mathbf{q})+i\Gamma-\frac{1}{2}\mathbf{\Omega}_{\mathbf{q}}\mathbb{\sigma}$. We note that only the zeroth diffraction order is accounted for in the far field, since the period of the structure is less than the wavelength. From Eq.~\eqref{El_full} it can be seen that the eigenfrequencies and eigenmodes of the whole structure are given by the condition $\mathrm{Det}\left[\hat{\alpha}_0(\mathbf{k+Q,\omega})^{-1}-\hat{G}_{\mathbf{k}}(0,0,\omega)\right]=0$, and the profile of the polariton mode can be restored from the kernel vector of this matrix. The kernel vector has two components of in-plane electric field for each diffraction order $\mathbf{Q}$. Therefore, we can image a polariton mode profile at specific $\mathbf{k}$ as an image on the reciprocal lattice where each node depicts a polarisation ellipse for a specific diffraction order. These plots are shown in Figs.~\ref{SI_fig2}(a,b) for the lower polariton mode at $\mathbf{k}=0$. The red and blue lines correspond to the two circular polarisations (helicities), while the violet color depicts linear polarisation. We can see that the helicity of the polariton near field $\eta$ satisfies $\eta(\mathbf{Q})=-\eta(-\mathbf{Q})$ due to the time reversal symmetry. We also note that while for the achiral structure (without the slant) the main diffraction orders are linearly polarised [Figs.~\ref{SI_fig2}(b)], for the chiral structure two main diffraction orders along $y$ axes have finite and opposite helicities [Figs.~\ref{SI_fig2}(a)]. The slant angle leads to the asymmetric scattering of $+1$ and $-1$ diffraction orders to the far field, which ultimately leads to the circularly polarised far-field emission.

We can calculate the far-field intensity from the electric field and average it over the random sources obtaining the final expression for the polarisation resolved PL intensity $I_{\alpha,\alpha'}$ 
\begin{align}
I_{\alpha'\alpha}(\mathbf{k})=& \gamma^{-2}R_{\alpha'\beta'}^{*\mathbf{Q'}} \rho_{\beta',\beta}(\mathbf{k+Q'},\omega)R_{\alpha\beta}^{\mathbf{Q'}}; \\ \quad\quad R_{\alpha\beta}^{\mathbf{Q'}}=&G_{\alpha\xi}^{0\mathbf{Q}}(\hat{\alpha}_{0}^{-1}-\hat{G})^{-1,\mathbf{QQ'}}_{\xi\beta} \label{I_fin}
\end{align}
Equation~\eqref{I_fin} relates the far-field emission intensity to the stationary distribution of the exciton pump rate $\rho$. We note that this distribution is qualitatively different from the thermal distribution since the system is out of equilibrium due to pump and radiative decay. Moreover, since the polaritons have negative effective mass, there can be no thermal equilibrium in the system. While precise determination of the distribution function would require the solution of a master equation for the density matrix accounting for the exciton-phonon and exciton-exciton scattering, we can adopt a generic expression for $\rho_{\alpha\beta}$:
\begin{align}
\rho_{\alpha\beta}(\mathbf{q},\omega)=\rho^{(0)}_{\alpha\beta} (1-e^{- q^2/\tilde{q}^2}), \label{eq:fit}
\end{align}
where $\rho^{(0)}$ is the polarisation matrix of the optical pump, and $\tilde{q}$ is the fitting parameter. The dependence in Eq.~\eqref{eq:fit} reflects the suppression of the pump to $q=0$ states and a relatively stronger pump to the high $q$ states lying outside the light cone. 

The calculated reflectivity and PL spectra based on our model for chiral (with the slant) and achiral (without the slant) geometries are presented in $\ref{SI_fig2}c-f$. The different underlying physics between reflectivity and PL spectra are seen in both cases, where the exciton response dominates the reflectivity spectrum, while the polariton emission dominates the PL spectrum. This shows that the intensity enhancement of the polariton PL is mainly due to the Brillouin zone folding. This effect  overcomes the thermalisation process, as discussed in the main text, and therefore is independent on chirality. However, as seen for the lower polariton, the chiral geometry induces asymmetry between the spectra produced by $\sigma^+$ and $\sigma^-$ polarised white and laser light, respectively. This behavior underlines the intrinsically chiral nature of the exciton polaritons forming in the regime of strong coupling between chiral BIC photons and excitons. 

\begin{figure}[htp]
\centering
\includegraphics[width=8.5 cm]{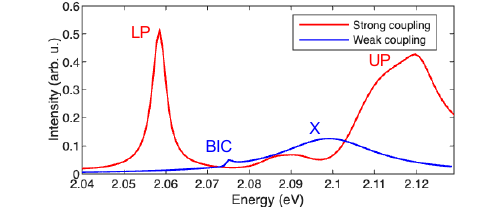}
\caption{{\bf {\color{black}Calculated absolute PL intensities for weakly and strongly coupled systems.}} {\color{black}The lower (LP) and upper (UP) polariton peaks in the strong coupling regime are shown in red  and the chiral BIC and exciton (X) peaks in the weak coupling regime are shown in blue.}} 
\label{SI_fig5_new}
\end{figure}

{\color{black}
To estimate the effects of strong coupling on the total PL intensity, we compare the calculated PL spectrum with that of a comparable system in the weak coupling regime. For the latter, we added 100 nm spacing between the monolayer and the metasurface, which substantially inhibited energy exchange interaction between the excitons and photons. The resulting spectra are shown in Figure \ref{SI_fig5_new}. In addition to the two-level repulsion, strong coupling significantly enhances overall PL emission from both the intrinsically chiral (lower polariton) state and the more exciton-like upper polariton state, as discussed in the main text.
}

{\bf Experiment:} 
\begin{figure}[htp]
\centering
\includegraphics[width=8.5 cm]{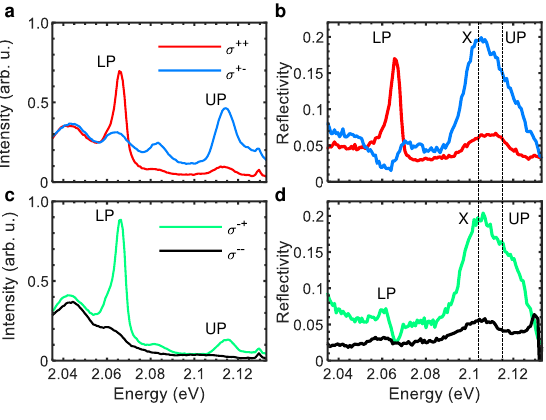}
\caption{{\bf Polarisation resolved reflectivity and PL spectra  of our structure.} {\bf a,b} Polarisation resolved photoluminescence and reflectivity spectra at $k=0$ using $\sigma^+$ polarised laser excitation or white light, respectively. {\bf c,d} Polarization resolved photoluminescence and reflectivity spectra at $k=0$ using $\sigma^-$ polarised laser excitation or white light, respectively.}
\label{SI_fig3}
\end{figure}
To probe polaritons and excitons in our system, we performed polarisation resolved PL and reflectivity measurements at normal incidence (see Fig. \ref{SI_fig3}a-d)). Therefore, we employed either a circularly polarised cw-laser ($\lambda=561 nm$) or circularly polarised white light. For $\sigma^+$ polarised light sources (see Fig. \ref{SI_fig3}a,b), the $\sigma^+$ polarised component of the emitted and reflected light  yields a strong lower polariton peak. The $\sigma^-$ component, on the contrary, stems mainly from the upper polariton in the PL and from the exciton in the reflectivity measurement. This behavior changes dramatically for $\sigma^-$ polarised light sources (see Fig. \ref{SI_fig3}c,d), where contributions of excitons and polaritons are all strongest for the $\sigma^+$ polarised component of the emitted and reflected light, and suppressed for the  $\sigma^-$ polarised component. This results from the interplay between intrinsically chiral states and rotating dipoles, combined with diffraction induced photoluminescence, as discussed in the main text. As the lower polariton possess robust $\sigma^+$-polarisation, independent on the polaritons of the pump, and is only receptive to $\sigma^+$-polarised whitelight, we conclude that the lower polariton is an intrinsically chiral light-matter hybrid state, inheriting the physics from the intrinsically chiral photonic BIC \cite{Chen2023}. As the polaritons have a matter (exciton) component, we also expect direct interactions with the environment, e.g., with phonons, which we probe by performing temperature dependent measurements.

\begin{figure}[htp]
\centering
\includegraphics[width=8.5 cm]{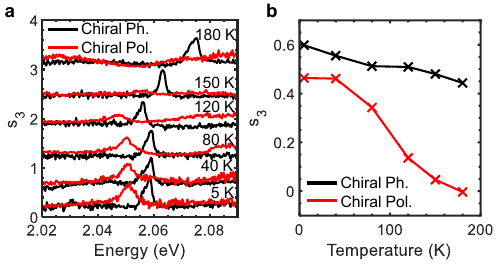}
\caption{{\color{black}{Temperature dependent measurements of the degree of circular polarisation.} {\bf a}. Waterfall plot of the s$_3$ spectra for (black) chiral photons in the bare metasurface and (red) polariton PL on the monolayer for various temperatures between 5 and 180 K. {\bf b} Peak amplitudes of the $s_3$ spectra shown in panel (a) as a function of temperature.}}
\label{SI_fig5}
\end{figure}

{\color{black}
As shown in Fig. 4g of the main text, the energy and s$_3$ value of the lower polariton decreases, and the linewidth increases with temperature. These trends can be explained by the stronger interaction with phonons at higher temperatures, leading to depolarisation, reduction of lifetime, and energy exchange. We test the robustness of our observation further by fabricating another, similar structure, and repeating the reflectivity and PL measurements on the bare metasurface and the metasurface/WS$_2$ heterostructure, respectively. Figure \ref{SI_fig5}a,b shows the s$_3$ spectra and peak s$_3$ values of the chiral photons and polaritons for a larger temperature range. In this sample, the chiral polariton PL diminishes at temperatures above 120 K, likely because of doping and less robust strong light-matter coupling compared to the sample discussed in the main text. However, between 5 K and 120 K, we observe polariton redshift and linewidth broadening with increasing temperature, similar to what we have observed in the sample discussed in the main text. Most strikingly, phonon-induced depolarisation of the chiral polariton causes a sharp drop of s$_3$ with temperature, while the photonic resonance in the metasurface remains chiral (see Fig. \ref{SI_fig5}b). Therefore, phonon-induced depolarisation and linewidth broadening is much stronger pronounced for the polariton compared to the photon, which can be explained by dominating temperature dependent phonon-polariton interactions via the excitonic component of polaritons.

\begin{figure}[htp]
\centering
\includegraphics[width=8.5 cm]{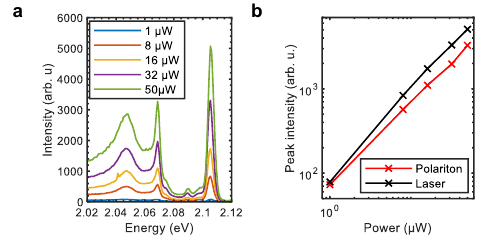}
\caption{{\color{black}{\bf Power dependent measurements of the lower polariton.} {\bf a} Power dependent PL spectra of lower polaritons (see text). {\bf b} Peak intensities of (black) the filtered laser at 2.13 eV and (red) the lower polariton.}}
\label{SI_fig7}
\end{figure}

To verify that our polaritons are in the thermal regime and behave like single particles, we performed power dependent measurements with 200 fs laser pulses tuned resonantly to the upper polariton state (see Fig. \ref{SI_fig7}). To suppress the pulsed laser peak in the measured spectra, we excited with horisontally polarised light, and measured only vertically polarised light in the detection path, using linear polarisers. When extracting the peak intensities for the lower polariton peak at $~\sim 2.07~\mathrm{eV}$ in Fig. \ref{SI_fig7}b, we can detect a slightly sub-linear input-output characteristics for the polaritons in comparison to the pump laser. Thus, the polaritons exhibit no nonlinear behavior and behave like single particles. This confirms that the polaritons are in the thermal regime within the shown range of average powers, and we can estimate their lifetime and temporal coherence from the linewidth measurement \cite{Wurdack2021}.

\begin{figure}[htp]
\centering
\includegraphics[width=8.5 cm]{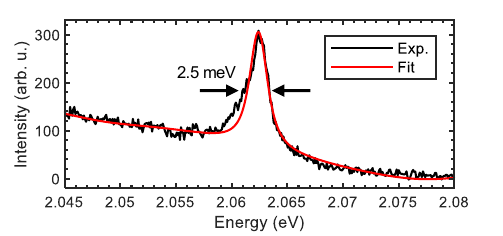}
\caption{{\color{black}High-resolution spectrum of the PL, with the chiral lower polariton state fitted with a Voigt profile.}}
\label{SI_fig6}
\end{figure}

To estimate the lifetime, we measured a high resolution spectrum of the chiral lower polariton, as presented in Fig. \ref{SI_fig6}. By fitting a Voigt profile to the spectrum, we can retrieve both homogeneous and inhomogeneous linewidth broadening ($\Delta E^H$ and $\Delta E^{IH}$, respectively), whereas  homogeneous broadening scales inversely with the lifetime. Thus, with the fitted linewidths of $\Delta E^{H}\approx2.5$ meV and $\Delta E^{IH}\approx 0.2$ meV, the polaritons have an estimated lifetime of $\tau = \hbar/\Delta E \approx 530~\mathrm{fs}$, which is consistent with the coherence times previously determined for confined WS$_2$ polariton states with low inhomogeneous broadening \cite{Wurdack2021}.

}


\bibliography{refs_supp.bib}